\documentclass[aps,prd,superscriptaddress,showpacs,preprint,amsmath,amssymb]{revtex4}
\usepackage{graphicx, bm}
\usepackage[usenames]{color}

\begin{document}

\draft
\title{Bounds on the non-standard $W^+W^-\gamma$ couplings at the LHeC and the FCC-he}

\author{M. K\"{o}ksal\footnote{mkoksal@cumhuriyet.edu.tr}}
\affiliation{\small Deparment of Optical Engineering, Sivas Cumhuriyet University, 58140, Sivas, Turkey.\\}

\author{A. A. Billur\footnote{abillur@cumhuriyet.edu.tr}}
\affiliation{\small Deparment of Physics, Sivas Cumhuriyet University, 58140, Sivas, Turkey.\\}

\author{ A. Guti\'errez-Rodr\'{\i}guez\footnote{alexgu@fisica.uaz.edu.mx}}
\affiliation{\small Facultad de F\'{\i}sica, Universidad Aut\'onoma de Zacatecas\\
         Apartado Postal C-580, 98060 Zacatecas, M\'exico.\\}

\author{ M. A. Hern\'andez-Ru\'{\i}z\footnote{mahernan@uaz.edu.mx}}
\affiliation{\small Unidad Acad\'emica de Ciencias Qu\'{\i}micas, Universidad Aut\'onoma de Zacatecas\\
         Apartado Postal C-585, 98060 Zacatecas, M\'exico.\\}

\date{\today}

\begin{abstract}

We examine the potential of the $e^-p \to e^-\gamma^*p \to e^-W^-q'X$ ($\gamma^*$ is Weizsacker-Willams photon) reaction
to probe the non-standard $W^+W^-\gamma$ couplings at the Large Hadron electron Collider (LHeC) and the Future Circular
Collider-hadron electron (FCC-he). We find $95\%$ confidence level bounds on the anomalous coupling $\Delta\kappa_\gamma$
and $\lambda_\gamma$ parameters in the view of effective Lagrangian approach with various values of the integrated luminosity.
We assume center-of-mass energies of the electron-proton system $\sqrt{s}= 1.30, 1.98, 7.07, 10\hspace{0.8mm}{\rm TeV}$ and
luminosities ${\cal L} = 10-1000 \hspace{0.8mm}{\rm fb^{-1}}$. The best limits obtained from the process
$e^-p \to e^-\gamma^*p \to e^-W^-q'X$ on the anomalous $\Delta\kappa_\gamma$ and $\lambda_\gamma$ couplings are
$\Delta\kappa_\gamma = |0.00069|$ and $\lambda_\gamma = [-0.0099, 0.0054]$. These bounds show that the process under
consideration is a good prospect for the searching of the non-standard $\Delta\kappa_\gamma$ and $\lambda_\gamma$ couplings
at the LHeC and the FCC-he. In addition, our results provide complementary information on other results for $\Delta\kappa_\gamma$
and $\lambda_\gamma$ couplings.

\end{abstract}


\vspace{5mm}

\maketitle

\section{Introduction}

The Standard Model (SM) of Elementary Particle Physics \cite{SM1,SM2,SM3} is extremely predictive and has been tested in numerous
aspects with impressive precision. The SM is a very powerful tool to predict the characteristics, the behavior and the interactions
of the elementary particles. Therefore, it is very important to measure particle properties and interactions in the most accurate way
possible to better understand the SM, refine it and test its global consistency. In this regard, the anomalous Triple-Gauge-Boson
Couplings (aTGC) and anomalous Quartic Gauge Boson Couplings (aQGC) of the $W^{\pm}$ boson: $W^+W^-\gamma$, $W^+W^-Z$, $WZ\gamma$,
$W\gamma\gamma$, $W^+W^-\gamma\gamma$, $W^+W^-Z\gamma$, $W^+W^-ZZ$ and $W^+W^-W^+W^-$ are a key element in the search
for the new physics beyond the SM (BSM), since any discrepancy of the measured value with respect to the predicted could reveal
new phenomena other than the SM.

In particular, the anomalous contribution to the $W^+W^-\gamma$ vertex, that is $\Delta\kappa_\gamma$ and $\lambda_\gamma$ parameters
have been studied for the ATLAS \cite{ATLAS}, CMS \cite{CMS}, CDF \cite{CDF}, D0 \cite{D0}, ALEP, DELPHI, L3, OPAL \cite{LEP} and TESLA
\cite{TESLA} experiments. For other experimental and phenomenological reports on $\Delta\kappa_\gamma$ and $\lambda_\gamma$, see Table I,
as well as Refs. \cite{Gutierrez,Baur0,Hagiwara,Hagiwara1,Nachtmann,Sahin0,Cakir,Ari,Atag,Atag1,Sahin,Papavassiliou,
Choudhury,Chapon,Ellis,Disha,Sahin5,Kumar}. Bounds on the aTGC are discussed in Subsection III-B.

These studies underline the importance to measure the anomalous $\Delta\kappa_\gamma$ and $\lambda_\gamma$ gauge couplings
in several different channels, contexts and colliders. For instance, in hadron-hadron, lepton-lepton and hadron-lepton
colliders, such as the Large Hadron Collider (LHC), the Compact Linear Collider (CLIC) and the Future Circular Collider
(FCC) at CERN. Furthermore, these colliders, operating in the $e^-\gamma$, $\gamma\gamma$, $e^-\gamma^*$, $\gamma^*p$
and $\gamma^*\gamma^*$ modes will be very useful. Some characteristics that distinguish to the future circular colliders
from the linear ones are the following: 1) Circular colliders can reach higher-luminosity, especially at low energies.
2) It could start operating as a factory of pairs of $Z$  and $W^\pm$ bosons.

With these arguments, we determined the production cross-section, as well as model-independent bounds for the non-standard
$W^+W^-\gamma$ couplings at the Large Hadron electron Collider (LHeC) and the Future Circular Collider-hadron electron
(FCC-he) \cite{FCChe,Fernandez,Fernandez1,Fernandez2,Huan,Acar}, through the $e^-p \to e^-\gamma^*p \to e^-W^-q'X$ reaction,
where $\gamma^*$ is Weizsacker-Willams photon and $X$ represents the proton remnants after deep inelastic scattering.
In this study, we use projections for runs at center-of-mass energies of $1.30, 1.98, 7.07 \hspace{0.8mm}{\rm TeV}$
and $10\hspace{0.8mm}{\rm TeV}$ and total integrated luminosities of $10, 30, 50, 70,100, 300, 500, 700$ and
$1000\hspace{0.8mm}{\rm fb^{-1}}$, respectively. In addition, to characterize possible deviations from the SM predictions
on the anomalous $\Delta\kappa_\gamma$ and $\lambda_\gamma$ couplings, we employ the effective Lagrange technique where the
SM is extended by a set of dimension-six operators.

The rest of this paper is organized as follows. In Sect. II, a brief review of the operators in effective Lagrangian is provided.
In Sect. III, we compute the total cross-section and derive bounds for the anomalous $\Delta\kappa_\gamma$ and
$\lambda_\gamma$ couplings at the LHeC and the FCC-he. In Sect. IV, we present our conclusions.

\begin{table}
\caption{Summary of experimental and phenomenological bounds at $95\%$ C.L. on the aTGC $\Delta\kappa_\gamma$ and $\lambda_\gamma$
from the present and future colliders.}
\begin{center}
\begin{tabular}{|c|c|c|c|c|}
\hline\hline
{\bf Model}             &    {$\Delta\kappa_\gamma$}  &  {$\lambda_\gamma$}          & {\bf C. L.}   &  {\bf Reference}\\
\hline
SM                      &           0                 &    0                         &               &   \cite{SM1,SM2,SM3}  \\
\hline
\hline
\hline
{\bf Experimental limit}      &  {$\Delta\kappa_\gamma$}    &  {$\lambda_\gamma$}          &  {\bf C. L.}  &  {\bf Reference}\\
\hline
ATLAS Collaboration              &    [-0.061, 0.064]          & [-0.013, 0.013]      & $95 \%$    & \cite{ATLAS} \\
\hline
CMS Collaboration                &    [-0.044, 0.063]          & [-0.011, 0.011]      & $95 \%$    & \cite{CMS} \\
\hline
CDF Collaboration                &    [-0.158, 0.255]          & [-0.034, 0.042]      & $95 \%$    & \cite{CDF} \\
\hline
D0 Collaboration                 &    [-0.158, 0.255]          & [-0.034, 0.042]      & $95 \%$    & \cite{D0} \\
\hline
ALEP, DELPHI, L3, OPAL           &    [-0.099, 0.066]          & [-0.059, 0.017]      & $95 \%$    & \cite{LEP} \\
\hline
\hline
\hline
{\bf Phenomenological limit}           &  {$\Delta\kappa_\gamma$}    &  {$\lambda_\gamma$}  &{\bf C. L.} &  {\bf Reference}\\
\hline
LHC                              &    [-0.182, 0.793]          & [-0.065, 0.065]  & $95 \%$    & \cite{LHC} \\
\hline
LHeC                              &    [[−0.0016, 0.0024]       & [-0.0040, 0.0043]  & $95 \%$    & \cite{LHeC} \\
\hline
CEPC                             &    [-0.00045, 0.00045]      & [-0.00033, 0.00033]  & $95 \%$    & \cite{CEPC} \\
\hline
ILC                              &    [-0.00037, 0.00037]      & [-0.00051, 0.00051]  & $95 \%$    & \cite{ILC} \\
\hline
CLIC                             &    [-0.00007, 0.00007]      & [-0.00004, 0.00004]  & $95 \%$    & \cite{Billur} \\
\hline\hline
\end{tabular}
\end{center}
\end{table}

\section{ A brief review of the non-standard $W^+W^-\gamma$ couplings}

Usually, deviations from the SM prediction are parameterized by an effective Lagrangian which contains, in addition
to the renormalizable part of the SM, a series of higher-dimensional effective operators suppressed by the scale of
new physics $\Lambda$.

In this context, in order to analyze the bounds on the non-standard $W^+W^-\gamma$ couplings $\Delta\kappa_\gamma$
and $\lambda_\gamma$ through the process $e^-p \to e^-\gamma^*p \to e^-W^-q'X$, we adopt the effective Lagrangian:

\begin{equation}
{\cal L}_{eff}={\cal L}^{(4)}_{SM} + \sum_i \frac{C^{(6)}_i}{\Lambda^2}{\cal O}^{(6)}_i + {\rm h.c.},
\end{equation}

\noindent where ${\cal L}^{(4)}_{SM}$ denotes the renormalizable SM Lagrangian and the non-SM part contains ${\cal O}^{(6)}_i$ the
gauge-invariant operators of mass dimension-six. The index $i$ runs over all operators of the given mass dimension. The mass scale
is set by $\Lambda$, and the coefficients $C_i$ are dimensionless parameters, which are determined once the full theory is known.

Therefore, effective Lagrangian relevant to our study on $\Delta\kappa_\gamma$ and $\lambda_\gamma$ is given by:

\begin{equation}
{\cal L}_{eff}=\frac{1}{\Lambda^2}\Bigl[C_W{\cal O}_W + C_B{\cal O}_B + C_{WWW}{\cal O}_{WWW} + \mbox{h.c.}\Bigr],
\end{equation}

\noindent where the ${\cal O}_W$, ${\cal O}_B$ and  ${\cal O}_{WWW}$ operators parameterize the non-standard $W^+W^-\gamma$
interactions:

\begin{eqnarray}
{\cal O}_W&=&\bigl( D_\mu\Phi \bigr)^{\dagger}\hat W^{\mu\nu}\bigl( D_\nu\Phi \bigr),\\
{\cal O}_B&=&\bigl( D_\mu\Phi \bigr)^{\dagger}\hat B^{\mu\nu}\bigl( D_\nu\Phi \bigr),\\
{\cal O}_{WWW}&=&Tr\bigl[\hat W^{\mu\nu}\hat W^\rho_\nu\hat W_{\mu\rho}\bigr].
\end{eqnarray}

\noindent From Eqs. (3)-(5), $D_\mu$ is the covariant derivative and $\Phi$ is the Higgs doublet field. $\hat B_{\mu\nu}$,
and $\hat W_{\mu\nu}$ are the $U(1)_Y$ and $SU(2)_L$ gauge field strength tensors, and the coefficients of these operators
$C_W/\Lambda^2$, $C_B/\Lambda^2$, and $C_{WWW}/\Lambda^2$, are zero in the SM.

With this methodology, the effective Lagrangian for the non-standard $W^+W^-\gamma$ couplings of electroweak  gauge bosons take the
form \cite{,Hagiwara,Gaemers}:

\begin{equation}
{\cal L}_{WW\gamma}=-ig_{WW\gamma} \Bigl[g^\gamma_1(W^\dagger_{\mu\nu} W^\mu A^\nu - W^{\mu\nu} W^\dagger_\mu A_\nu )
+\kappa_\gamma W^\dagger_{\mu} W_\nu A^{\mu\nu} + \frac{\lambda_\gamma}{M^2_W} W^\dagger_{\rho\mu} W^\mu_\nu A^{\nu\rho} \Bigr],
\end{equation}

\noindent where $g_{WW\gamma}=e$, $V_{\mu\nu}=\partial_\mu V_\nu -\partial_\nu V_\mu$ with $V_\mu= W_\mu, A_\mu$. The couplings
$g^\gamma_1$, $\kappa_\gamma$ and $\lambda_\gamma$ are CP-preserving, and in the SM, $g^\gamma_1=\kappa_\gamma=1$ and $\lambda_\lambda=0$
at the tree level.

The three operators of dimension-six, given by Eq. (2) are related to the anomalous triple gauge boson couplings via \cite{Rujula,Hagiwara1,Hagiwara2}:

\begin{equation}
\kappa_\gamma= 1+ \Delta\kappa_\gamma,
\end{equation}

\noindent with

\begin{eqnarray}
\Delta\kappa_\gamma&=&C_W + C_B,\\
\lambda_\gamma&=&C_{WWW}.
\end{eqnarray}

From the effective Lagrangian given in Eq. (6), the Feynman rule for the anomalous $W^+W^-\gamma$ vertex function most general
CP-conserving and that is consistent with gauge and Lorentz invariance of the SM is given by \cite{Hagiwara}:

\begin{eqnarray}
\Gamma^{WW\gamma}_{\mu\nu\rho}&=&e\Bigl[g_{\mu\nu}(p_1-p_2)_\rho + g_{\nu\rho}(p_2-p_3)_\mu + g_{\rho\mu}(p_3-p_1)_\nu
+\Delta\kappa_\gamma \Bigl( g_{\rho\mu}p_{3\nu}- g_{\nu\rho}p_{3\mu}\Bigr )  \nonumber\\
&+&\frac{\lambda_\gamma}{M^2_W}\Bigl(p_{1\rho}p_{2\mu}p_{3\nu} - p_{1\nu}p_{2\rho}p_{3\mu}
- g_{\mu\nu}(p_2\cdot p_3 p_{1\rho}-p_3\cdot p_1 p_{2\rho}) \nonumber\\
&-& g_{\nu\rho}(p_3\cdot p_1 p_{2\mu}-p_1\cdot p_2 p_{3\mu})
- g_{\mu\rho}(p_1\cdot p_2 p_{3\nu}-p_2\cdot p_3 p_{1\nu}) \Bigr)\Bigr],
\end{eqnarray}

\noindent where $p_1$ represents the momentum of the photon and $p_2$ and $p_3$ symbolize the momenta of $W^\pm$ bosons.
The first three terms in Eq. (10) corresponds to the SM couplings, while the terms with $\Delta\kappa_\gamma$ and
$\lambda_\gamma$ give rise to the anomalous triple gauge boson couplings.

Several searches on these anomalous $\Delta\kappa_\gamma$ and $\lambda_\gamma$ couplings of $W^+W^-\gamma$ vertex were
performed by the LEP, Tevatron and LHC experiments, as shown in Table I.

\section{Cross-section of the process $e^-p \to e^-\gamma^*p \to e^-W^- q'X$ and bounds on the anomalous
$\Delta\kappa_\gamma$ and $\lambda_\gamma$ couplings at the LHeC and the FCC-he}

\subsection{Cross-section of the process $e^-p \to e^-\gamma^*p \to e^-W^- q'X$ at the LHeC and the FCC-he}

At the LHeC and the FCC-he, the aTGC can be directly probed via single-$W^\pm$ production.
Therefore, in this work we focus on the process $e^-p \to e^-\gamma^*p \to e^-W^- q'X$, as well as with its corresponding
sub-processes (see Figs. 1 and 2):

\begin{eqnarray}
\gamma^* u &\to& W^+d \to l^+ \nu_l d, \\
\gamma^* u &\to& W^+s \to l^+ \nu_l s,
\end{eqnarray}

\begin{eqnarray}
\gamma^* c &\to& W^+s \to l^+ \nu_l s,  \\
\gamma^* c &\to& W^+d \to l^+ \nu_l d,
\end{eqnarray}

\begin{eqnarray}
\gamma^* \bar d &\to& W^+\bar u \to l^+ \nu_l \bar u,  \\
\gamma^* \bar d &\to& W^+\bar c \to l^+ \nu_l \bar c,
\end{eqnarray}

\begin{eqnarray}
\gamma^* \bar s &\to& W^+\bar c \to l^+ \nu_l \bar c,  \\
\gamma^* \bar s &\to& W^+\bar u \to l^+ \nu_l \bar u,
\end{eqnarray}

\begin{eqnarray}
\gamma^* \bar b &\to& W^+\bar t \to l^+ \nu_l \bar t,
\end{eqnarray}

\noindent where $l^+=e^+, \mu^+$; $\nu_l=\nu_e,\nu_\mu$ and $\gamma^*$ is the Weizsacker-Williams photon. It is worth noting
that in Eqs. (11)-(19), we take into account interactions between different flavors of the quarks due to off-diagonal elements
of the Cabibbo-Kobayashi-Maskawa matrix.

In this type of $\gamma^*p$ interactions the emitted quasi-real photon $\gamma^*$ is scattered with small angles from the beam pipe
of $e^-$ \cite{Ginzburg,Ginzburg1,Brodsky,Budnev,Terazawa,Yang}. As these photons have low virtuality, they are almost on the
mass shell. These processes can be described by the Equivalent Photon Approximation (EPA) \cite{Budnev,Baur1,Piotrzkowski}, using
the Weizsacker-Williams Approximation (WWA). The EPA has a lot of advantages such as providing the skill to reach crude numerical
predictions via simple formulae. Furthermore, it may principally ease the experimental analysis because it enables one to directly
achieve a rough cross-section for $\gamma^{*}p \to X$ process via the examination of the main process $e^{-}p\rightarrow e^{-} X p$
where X represents objects produced in the final state. The production of high mass objects is particularly interesting at the $e^-p$
colliders and the production rate of massive objects is limited by the photon luminosity at high invariant mass while the $\gamma^{*}p$
process at the $e^-p$ colliders arises from quasi-real photon emitted from the incoming beams. However, using the EPA, new physics
searches BSM have been theoretically and experimentally examined at the LEP, the Tevatron and the LHC \cite{Abulencia,Aaltonen1,Aaltonen2,Chatrchyan1,Chatrchyan2,Abazov,Chatrchyan3,Inan,Inan1,Inan2,Sahin1,Atag2,Sahin2,Sahin4,Senol,
Senol1,Fichet,Sun,Sun1,Sun2,Senol2}.

In the EPA, the energy spectrum of the photons emitted from electron beam is given by the following analytical formula \cite{Belyaev,Budnev}:

\begin{eqnarray}
f_{\gamma^{*}_e}(x_{1})=\frac{\alpha}{\pi E_{e}}\Bigl\{[\frac{1-x_{1}+x_{1}^{2}/2}{x_{1}}]log(\frac{Q_{max}^{2}}{Q_{min}^{2}})-\frac{m_{e}^{2}x_{1}}{Q_{min}^{2}}
(1-\frac{Q_{min}^{2}}{Q_{max}^{2}})-\frac{1}{x_{1}}[1-\frac{x_{1}}{2}]^{2}
log\Big(\frac{x_{1}^{2}E_{e}^{2}+Q_{max}^{2}}{x_{1}^{2}E_{e}^{2}+Q_{min}^{2}}\Big)\Big\}, \nonumber \\
\end{eqnarray}

\noindent where $x_1=E_{\gamma^*_e}/E_{e}$ is the energy fraction transferred from the electron to the photon and $Q^2_{\rm max}=2$ $\rm GeV^2$
is maximum virtuality of the photon. The minimum value of the $Q^2_{\rm min}$ is given by:

\begin{eqnarray}
Q_{\rm min}^{2}=\frac{m_{e}^{2}x_{1}^{2}}{1-x_{1}}.
\end{eqnarray}

Therefore, the total cross-section of the $e^-p \to e^-\gamma^*p \to e^-W^- q'X$ reaction is determined through:

\begin{eqnarray}
\sigma(e^-p \to e^-\gamma^*p \to e^-W^- q'X)=\int f_{\gamma^{*}_e}(x_{1}){\hat\sigma}(\gamma^*q \to W^-q') dx_{1}.
\end{eqnarray}

For the numerical evaluation of the total cross-section, we have used CTEQ6L1 \cite{Jonathan} for the parton distribution functions
and the $W^+W^-\gamma$ vertex is embedded in CalcHEP package \cite{Belyaev} together with the energy spectrum of the photon.
In all calculations in this paper, we assume that the center-of-mass energies of the electron-proton system are 1.30, 1.98, 7.07
and 10 TeV, respectively.

In Figs. 3 and 4, we plot the total cross sections of the $e^-p \to e^-\gamma^*p \to e^-W^- q'X$ process. We apply the following
basic cuts to reduce the background and optimize the signal:

\begin{eqnarray}
-2.5<&Y(e^+)&<2.5, \nonumber \\
-2.5<&Y(\mu^+)&<2.5.
\end{eqnarray}

\begin{eqnarray}
-5<&Y(d)&<5,  \nonumber\\
-5<&Y(s)&<5,  \nonumber\\
-5<&Y(\bar u)&<5,   \\
-5<&Y(\bar c)&<5. \nonumber
\end{eqnarray}

\begin{eqnarray}
T(e^+)&>&20 \hspace{0.8mm}{\rm GeV},  \nonumber\\
T(\mu^+)&>&20 \hspace{0.8mm} {\rm GeV}.
\end{eqnarray}

\begin{eqnarray}
T(\nu_e)&>&15 \hspace{0.8mm}{\rm GeV},  \nonumber\\
T(\nu_\mu)&>&15 \hspace{0.8mm}{\rm GeV}.
\end{eqnarray}

\begin{eqnarray}
T(d)&>&20 \hspace{0.8mm}{\rm GeV},  \nonumber\\
T(s)&>&20 \hspace{0.8mm}{\rm GeV}.
\end{eqnarray}

\begin{eqnarray}
T(\bar u)&>&20 \hspace{0.8mm}{\rm GeV},  \nonumber\\
T(\bar c)&>&20 \hspace{0.8mm}{\rm GeV}.
\end{eqnarray}

These graphs describe the behavior of the total cross-section $\sigma(\Delta\kappa_\gamma, \lambda_\gamma, \sqrt{s})$
in the context of effective Lagrangians, and as a function of the anomalous parameters $\Delta\kappa_\gamma/\lambda_\gamma $,
maintaining fixed the center-of-mass energy of the collider, which is assumed for values of $\sqrt{s}=1.30, 1.98, 7.07$ and 10 TeV.
It is observed that the total cross-section $\sigma(\Delta\kappa_\gamma, \lambda_\gamma, \sqrt{s})$ increases to a value of the order of
$\sim$ 90 pb (70 pb), on a scale of center-of-mass energies of 1.30 up to 10 TeV. From this analysis it is concluded that
the total cross-section to production single of vector bosons $W^-$ depends significantly on the center-of-mass energy of the collider,
as well as of the anomalous parameters $\Delta\kappa_\gamma$ and $\lambda_\gamma$, which means that the effective area of the
collision increases for very high energy ranges. In summary, the aTGC and the parameters of the colliders both tend to increase
the cross-section of the single $W^-$ gauge boson production.

\subsection{Bounds on the non-standard $\Delta\kappa_\gamma$  and $\lambda_\gamma$ couplings at the LHeC and the FCC-he}

A model-independent fit procedure to determine the precision of a quantity is based on the construction of a $\chi^2$ function
from all observables. Thus, the $\chi^2$ function to obtain our bounds on the non-standard $\Delta\kappa_\gamma$ and $\lambda_\gamma$
couplings at the $95\%$ Confidence Level (C.L.) can be expressed as \cite{Billur1,Mary,Koksal,Gutierrez,Koksal1,Gutierrez1}:

\begin{equation}
\chi^2(\Delta\kappa_\gamma, \lambda_\gamma )=\Biggl(\frac{\sigma_{SM}-\sigma_{NP}(\sqrt{s}, \Delta\kappa_\gamma, \lambda_\gamma)}{\sigma_{SM}\sqrt{(\delta_{st})^2+(\delta_{sys})^2}}\Biggr)^2.
\end{equation}

\noindent $\sigma_{SM}$ is the SM cross-section and $\sigma_{NP}(\sqrt{s}, \Delta\kappa_\gamma, \lambda_\gamma)$ is the
cross-section containing both the non-standard and SM contributions. $\delta_{st}=\frac{1}{\sqrt{N_{SM}}}$ is the statistical
error and $\delta_{sys}$ is the systematic error. The number of events is given by $N_{SM}={\cal L}_{int}\times \sigma_{SM}$.

\begin{table}[!ht]
\caption{The expected $95\%$ C.L. bounds for the anomalous couplings $\Delta\kappa_\gamma$
and $\lambda_\gamma$, through the process $ e^- p \rightarrow e^- \gamma^* p \rightarrow e^- W^- q\,' X $
for $\sqrt{s}=1.30, 1.98, 7.07, 10$ TeV and ${\cal L}=10, 30, 50, 70, 100, 300, 500, 700, 1000$ $\rm fb^{-1}$
at the LHeC and the FCC-he. The bounds for each parameter are calculated while fixing the other parameters to zero.}
\begin{center}
\begin{tabular}{|c|c|c|c|}
\hline\hline
\cline{1-4}
aTGC & ${\cal L} \, (\rm fb^{-1})$  & \hspace{0.8cm} $\sqrt{s}=$ 1.30 TeV \hspace{0.8cm} & \hspace{0.8cm} $\sqrt{s}=$ 1.98 TeV \hspace{0.8cm} \\
\hline
\hline
\cline{1-4}
 & 10  &  [-0.0290, 0.0283] & [-0.0195, 0.0191] \\
 & 30  &  [-0.0166, 0.0164] & [-0.0112, 0.0111] \\
\hspace{0.2cm} $\Delta\kappa_\gamma$ \hspace{0.2cm}
 & 50  &  [-0.0128, 0.0127] & [-0.0087, 0.0086] \\
 & 70  &  [-0.0108, 0.0107] & [-0.0073, 0.0072] \\
 & 100 &  [-0.0091, 0.0090] & [-0.0061, 0.0061] \\
\hline
 & 10  &  [-0.1783, 0.1280] & [-0.1122, 0.0822] \\
 & 30  &  [-0.1427, 0.0923] & [-0.0895, 0.0595] \\
$\lambda_\gamma$
 & 50  &  [-0.1293, 0.0789] & [-0.0810, 0.0509] \\
 & 70  &  [-0.1214, 0.0710] & [-0.0759, 0.0459] \\
 & 100 &  [-0.1137, 0.0634] & [-0.0710, 0.0410] \\
\hline
\hline
\cline{1-4}
aTGC & ${\cal L} \, (fb^{-1})$ & \hspace{0.8cm} $\sqrt{s}=$ 7.07 TeV \hspace{0.8cm} & \hspace{0.8cm} $\sqrt{s}=$ 10 TeV \hspace{0.8cm} \\
\hline
\hline
\cline{1-4}
 & 100  &  [-0.0027, 0.0027] & [-0.00210, 0.00210]  \\
 & 300  &  [-0.0015, 0.0015] & [-0.00120, 0.00120]  \\
\hspace{0.2cm}  $\Delta \kappa_\gamma$ \hspace{0.2cm}
 & 500  &  [-0.0012, 0.0012] & [-0.00097, 0.00097]  \\
 & 700  &  [-0.0010, 0.0010] & [-0.00082, 0.00082]  \\
 & 1000 &  [-0.0008, 0.0008] & [-0.00069, 0.00069]  \\
\hline
 & 100  &  [-0.0217, 0.0153] & [-0.01540, 0.01100]  \\
 & 300  &  [-0.0174, 0.0110] & [-0.01240, 0.00790]  \\
$\lambda_\gamma$
 & 500  &  [-0.0158, 0.0094] & [-0.01120, 0.00680]  \\
 & 700  &  [-0.0148, 0.0084] & [-0.01050, 0.00610]  \\
 & 1000 &  [-0.0139, 0.0075] & [-0.00990, 0.00540]  \\
\hline\hline
\end{tabular}
\end{center}
\end{table}

The calculated bounds at $95\%$ C.L. for the aTGC $\Delta\kappa_\gamma$ and $\lambda_\gamma$ is shown in Table II.
We have used that only one of the anomalous coupling is non zero at any given time, while the other anomalous
coupling is taken to zero. The bounds are computed and displayed separately for $\sqrt{s}=1.30, 1.98, 7.07, 10$ TeV
and ${\cal L}=10, 30, 50, 70, 100, 300, 500, 700, 1000$ $\rm fb^{-1}$.

The $\sqrt{s}=10$ TeV and ${\cal L}=1000$ $\rm fb^{-1}$ selection has significantly better sensitivity to the aTGC $\Delta\kappa_\gamma$
and $\lambda_\gamma$:

\begin{eqnarray}
\Delta\kappa_\gamma &=& |0.00069|, \hspace{3mm}   \mbox{$95\%$ C.L.}, \nonumber\\
\lambda_\gamma &=& [-0.0099, 0.0054], \hspace{3mm}   \mbox{$95\%$ C.L.}.
\end{eqnarray}

Our bounds on anomalous $\Delta\kappa_\gamma$ and $\lambda_\gamma$ couplings compare favorably with those reported by the
ATLAS \cite{ATLAS}, CMS \cite{CMS}, CDF \cite{CDF}, D0 \cite{D0}, ALEP, DELPHI, L3, OPAL \cite{LEP} and TESLA \cite{TESLA}
experiments (see Table I, Ref. \cite{Billur}). In addition, our bounds on $\Delta\kappa_\gamma$ and $\lambda_\gamma$ are
competitive with the phenomenological limits obtained by the LHC \cite{LHC}, the LHeC \cite{LHeC}, the ILC \cite{ILC},
the CEPC \cite{CEPC} and the CLIC \cite{Billur}, as well as those of Refs. \cite{Gutierrez,Baur0,Hagiwara,Hagiwara1,
Nachtmann,Sahin0,Cakir,Ari,Atag,Atag1,Sahin,Papavassiliou,Choudhury,Chapon,Ellis,Disha,Sahin5,Kumar}.

\section{Conclusions}

In conclusion, we studied and derived possible bounds on the aTGC $\Delta\kappa_\gamma$ and $\lambda_\gamma$ with the
$e^-p \to e^-\gamma^*p \to e^-W^-q'X$ reaction, where $\gamma^*$ is Weizsacker-Willams photon, together with the total
cross-section using ${\cal L}=10-1000$ $\rm fb^{-1}$ of $e^-p$ collisions and $\sqrt{s}=1.30, 1.98, 7.07, 10$ TeV at
the LHeC and the FCC-he.

The bounds obtained from our analysis on the $\Delta\kappa_\gamma$ and $\lambda_\gamma$ parameters are comparable to the previous
most stringent limits from other single-boson and diboson analyses. Furthermore, our results indicate (see Figs. 3-4, and Table II)
that the $e^-p \to e^-\gamma^*p \to e^-W^-q'X$ process is potentially suitable for probe the non-standard $W^+W^−\gamma$ couplings
at the LHeC and the FCC-he, and with cleaner environments, compared to those for the case of hadron colliders.

The topic is interesting and under-explored, so it is of great importance to have theoretical and phenomenological interest to motivate
experimental collaborations to measure this very interesting sector of QED. A prominent advantage of the $e^-p \to e^-\gamma^*p \to e^-W^-q'X$
process is that it isolates anomalous $W^+W^-\gamma$ couplings. It allows to study $W^+W^-\gamma$ couplings independent from $W^+W^-Z$ as well
as from $W^+W^-\gamma\gamma$.

\vspace{1cm}

\begin{center}
{\bf Acknowledgments}
\end{center}

A. G. R. and M. A. H. R. thank SNI and PROFEXCE (M\'exico).

\vspace{2cm}


\pagebreak

\begin{figure}[t]
\centerline{\scalebox{0.8}{\includegraphics{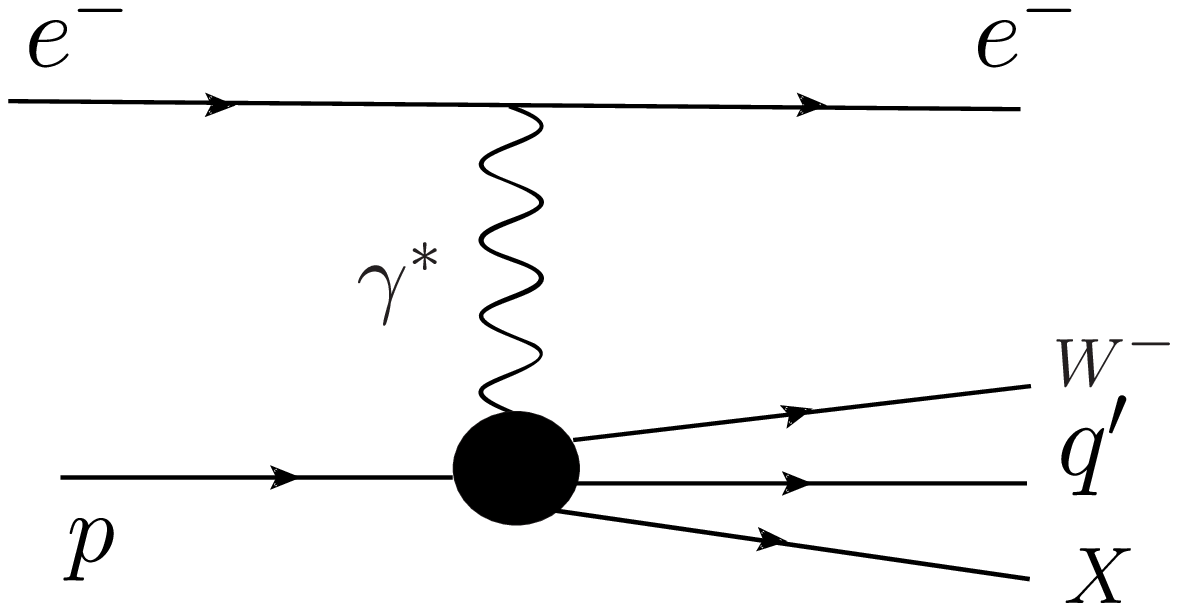}}}
\caption{ \label{fig:gamma1} A schematic diagram for the process
$e^-p \to e^-\gamma^*p \to e^-W^-q'X$.}
\label{Fig.1}
\end{figure}

\begin{figure}[t]
\centerline{\scalebox{0.8}{\includegraphics{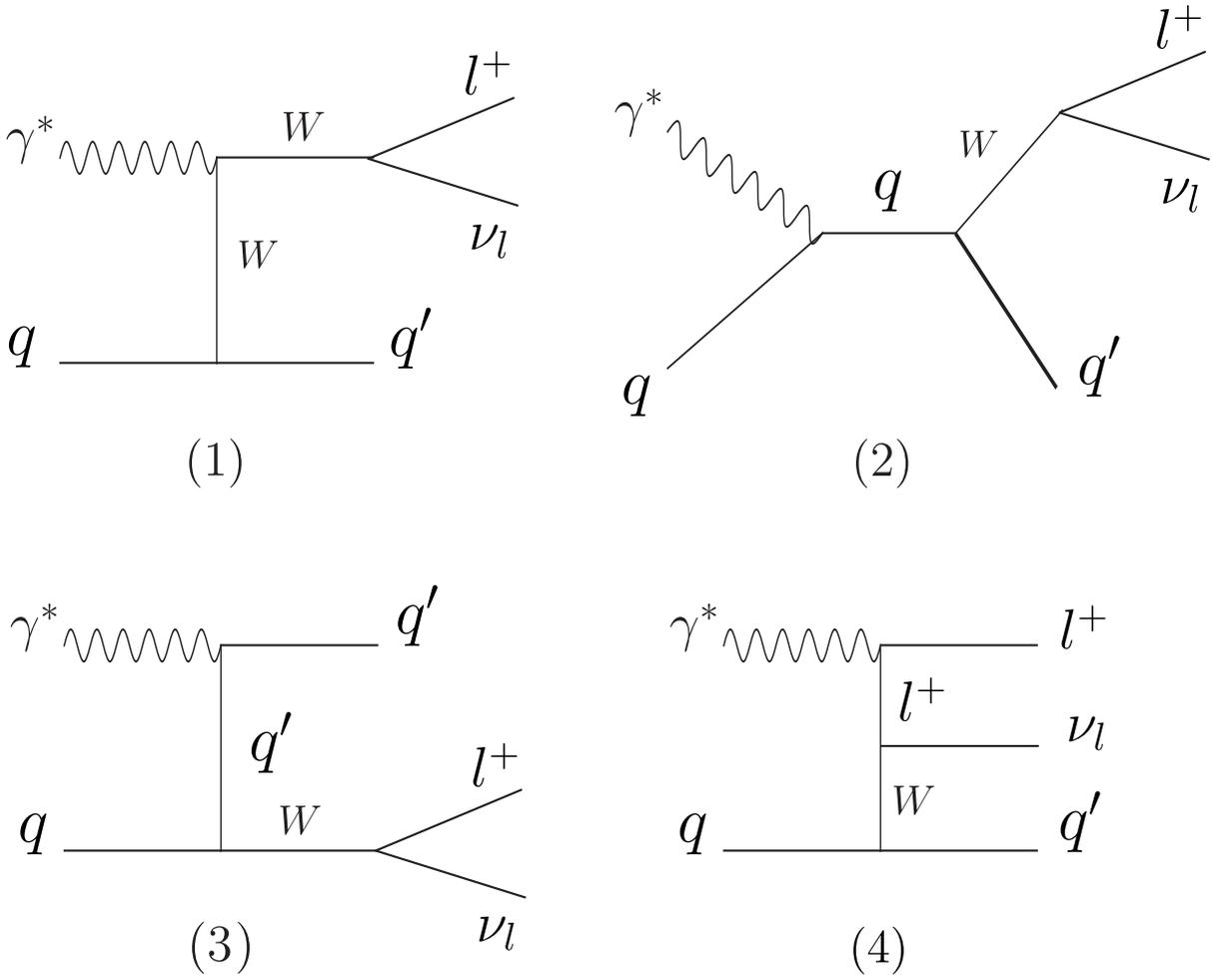}}}
\caption{ \label{fig:gamma2} Feynman diagrams contributing to the sub-process
$\gamma^*q \to W^-q'$.}
\label{Fig.2}
\end{figure}

\begin{figure}[t]
\centerline{\scalebox{1.4}{\includegraphics{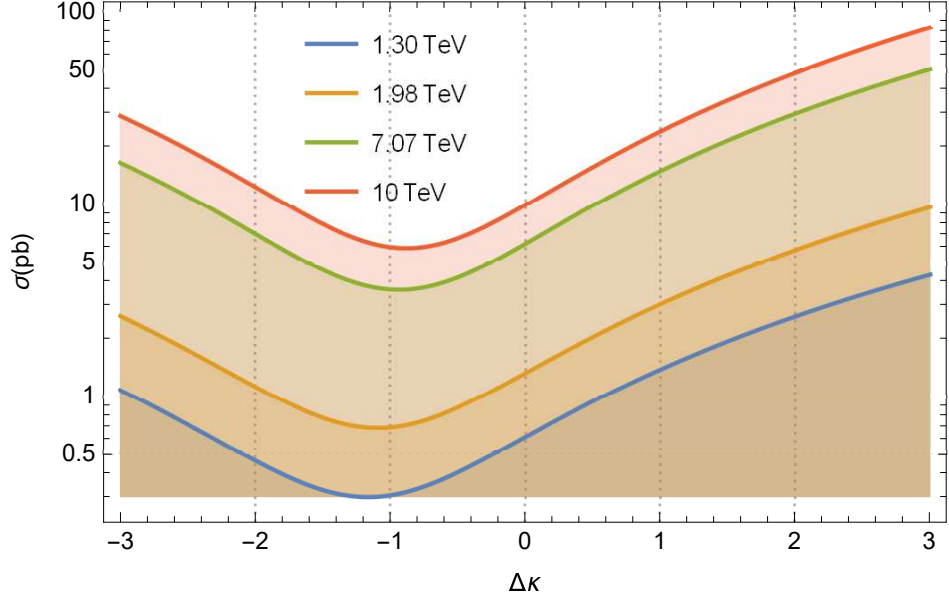}}}
\caption{ The total cross-sections of the process
$e^-p \to e^-\gamma^* p \to \nu_e W^- p$ as a function of $\Delta \kappa_\gamma$
for center-of-mass energies of $\sqrt{s}=1.30, 1.98, 7.07, 10\hspace{0.8mm}{\rm TeV}$ at the LHeC and the FCC-he.}
\label{Fig.3}
\end{figure}

\begin{figure}[t]
\centerline{\scalebox{1.4}{\includegraphics{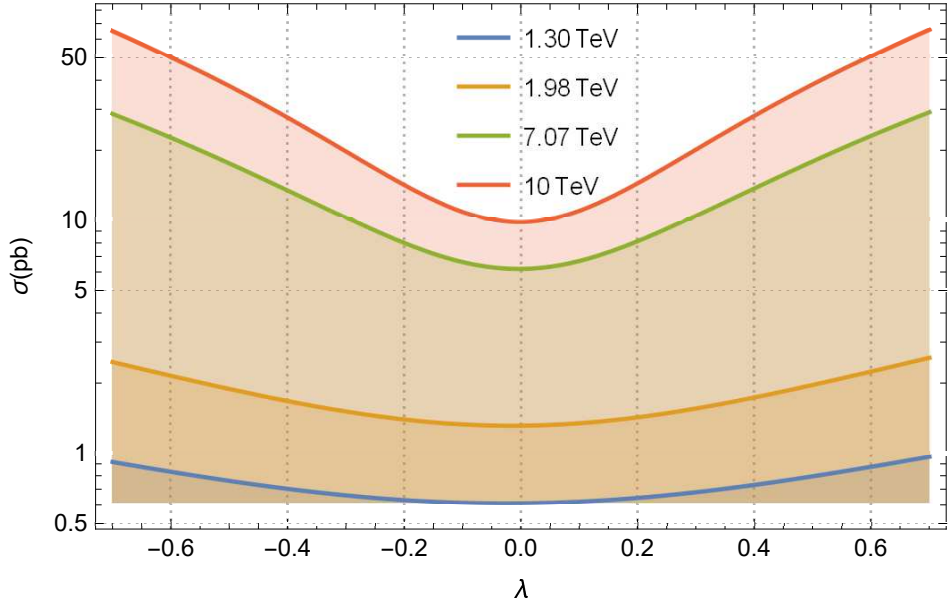}}}
\caption{ Same as in Fig. 3, but for $\lambda_\gamma$.}
\label{Fig.4}
\end{figure}

\end{document}